\documentclass[%
 reprint,
 amsmath,amssymb,
 aps,
]{revtex4-1}
\def\bg{\begin{equation}}
\def\nd{\end{equation}}
\newcommand{\be}{\begin{equation}} 
\newcommand{\ee}{\end{equation}} 
\newcommand{\bea}{\begin{eqnarray}} 
\newcommand{\eea}{\end{eqnarray}}

\usepackage{graphicx}% Include figure files
\usepackage{dcolumn}% Align table columns on decimal point
\usepackage{bm}% bold math
\begin{document}

\preprint{APS/123-QED}

\title{Vector Meson Spectrum from top-down Holographic QCD}% Force line breaks with \\

 \author{Mohammed Mia}%
\email{mmia@purdue.edu}
\affiliation{Department of Physics and Astronomy, Purdue University, 525 Northwestern Avenue, West Lafayette, IN 47907, USA}
\author{Keshav Dasgupta}
\email{keshavhep.physics.mcgill.ca}
 \affiliation{Department of Physics, McGill University, 3600 University Street, Montreal, QC, H3A 2T8, Canada}
\author{Charles Gale}
\email{gale@physics.mcgill.ca}
 \affiliation{Department of Physics, McGill University, 3600 University Street, Montreal, QC, H3A 2T8, Canada}
 \author{Michael Richard}
\email{michael.richard@mail.mcgill.ca}
 \affiliation{Department of Physics, McGill University, 3600 University Street, Montreal, QC, H3A 2T8, Canada}
\author{Olivier Trottier} 
\email{olivier.trottier@mail.mcgill.ca}
\affiliation{Department of Physics, McGill University, 3600 University Street, Montreal, QC, H3A 2T8, Canada}

\date{\today}% It is always \today, today,
             %  but any date may be explicitly specified

\begin{abstract}
We elaborate on the brane configuration that gives rise to a QCD-like gauge theory that confines at low energies and 
becomes scale invariant at the highest energies. In the limit where the rank of the gauge group is large, a gravitational
description emerges. For the confined phase, we obtain a vector meson 
spectrum and demonstrate how certain choice of parameters can lead to quantitative agreement with empirical data.

\end{abstract}

\maketitle

%\tableofcontents

\section{\label{Intro}Introduction}

Quantum Chromo-Dynamics (QCD) is the accepted gauge theory of the strong interaction, and its degrees of freedom are the fermionic quarks (and anti-quarks) and the bosonic gluons. The fact that the gluon gauge fields admit self-interaction creates a scale-dependent coupling: at short distances, smaller than that of a proton,  a quark-antiquark pair experiences a Coulomb-like potential while at larger distances the pair binds in a field configuration that is like that of a flux tube. In the regime where the coupling between partons becomes large, calculations of the low energy properties of QCD have traditionally relied either on large numerical simulation of the theory, discretized on a space-time lattice (``lattice QCD''), or on the use of effective theories based on some of the fundamental symmetries of QCD. However, a new class of approaches that have generated great interest is that based on the concept of holographic duality as expressed through the famous AdS/CFT correspondence \cite{Maldacena:1997re}. A powerful feature of such dualities is that even though the field theory sector might be strongly coupled, its gravity dual can be treated in perturbation theory. This opens the tantalizing prospect of being able to treat systems interacting through low-energy QCD analytically, at least to some extent. Theories that attempt to realize this potential are globally labelled AdS/QCD.  Even through no formal gravity dual of QCD is known, several models use holographic techniques to arrive at ``QCD-like'' field theories, at least to explain the infra-red (IR) dynamics. 
One important example being the
Klebanov-Strassler model \cite{Klebanov:2000hb}.
Bottom-up approaches are phenomenologically based and are not as such compelled by the rules of string theory. Top-down approaches start with gauge theory arising from open strings ending on branes and then study the dual closed string sector described by classical gravity. Such an approach, that in fact covers the regime from low to high energies, is used in this work; the next section describes this top-down model. Then, we construct a pseudo-QCD action from our gravity dual, and calculate the vector meson mass spectrum. We compare our results with those obtained using the Sakai-Sugimoto model, and also with experimentally measured masses. We end with a conclusion. 

\section{\label{BCon}Brane Configuration}

We start with coincident $N$ Dirichlet three-branes ($D3$ branes) and $M$ Dirichlet five-branes ($D5$ branes) at the tip of a cone [See Figure 1] and add $M$ anti-five
branes separated from each other and  from the $D3/D5$ branes [See Figure 1]. To obtain this separation, we must blow up one of the
$S^2$'s at the tip and give it a finite size. 
 The separation gives masses $\Lambda_0$ to the $D5/\bar{D}5$ strings and at scales $\Lambda<\Lambda_0$, the
 gauge group is $SU(N+M)\times SU(N)\times U(1)^M$ where the additional $U(1)$ groups arise from the massless strings ending 
 on the 
 $\bar{D}5$-branes spread above the equator of the $S^2$.  The $SU(M+N)$ sector has $2N$
effective flavors while the $SU(N)$ sector has $2(N+M)$ effective
flavors thus it is dual to the $SU(N-M)\times SU(N)$ gauge theory
under a Seiberg duality. Under a series of such dualities, which is
called cascading, at the far IR region the gauge theory can be
described by $SU(M)\times SU(K)$ group, where $N=lM+K$, $l$, $0\leq
K<M$ are positive integers. Now the number of `actual' D3 branes $N$
is no longer the relevant quantity, rather $N\pm pM$
where $p$ is an integer describes the D3 brane charge. We take $K=0$ in all our analysis, so at
the bottom of the cascade,
 we are left with  ${\cal N}=1$ SUSY $SU(M)$ strongly coupled
 gauge theory which looks very much like strongly coupled  SUSY QCD.
 
 \begin{figure}[htb]\label{gauge}
       \begin{center}
\includegraphics[height=5cm]{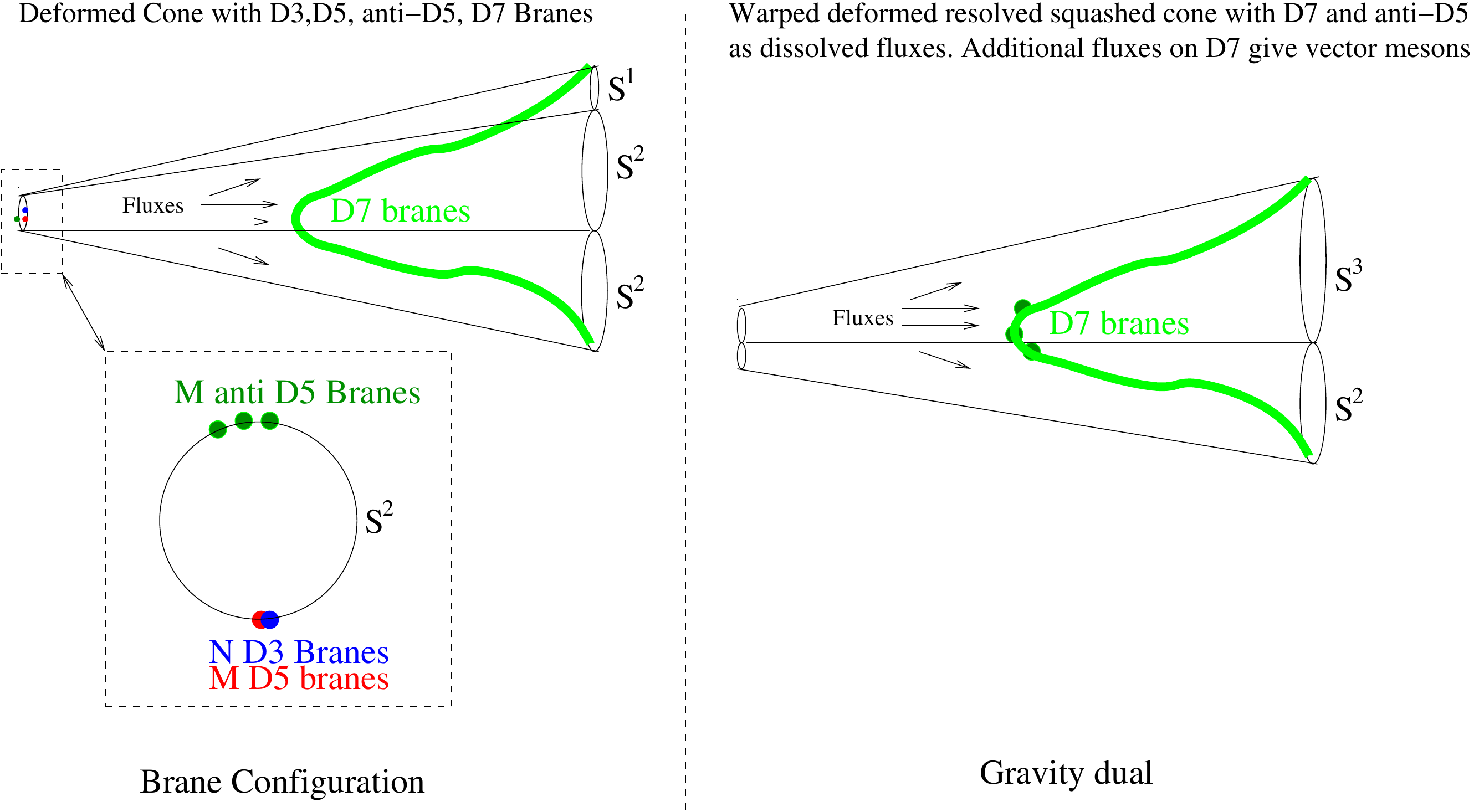}
        \caption{Brane configuration and the dual gravity in the extremal limit for a UV regular theory. The anti-branes should be thought of as spread above the equator of the resolved sphere although the branes are all localised at the south pole of the 
        sphere. The manifolds appearing on both sides of the duality are in general non-K\"ahler manifolds although in the limit of
        vanishing resolution and squashing they become K\"ahler Calabi-Yau spaces.}
       \end{center}
        \end{figure} 
At high energies $\Lambda>>\Lambda_0$, $D5/\bar{D}5$ strings are excited and we have $SU(N+M)\times SU(N+M)$
 gauge theory. Essentially $M$ pairs of $D5/\bar{D}5$ branes with fluxes are equivalent to $M$ number of  $D3$ branes and hence
 they contribute an additional $M$ units of $D3$ charge, resulting in $SU(N+M)\times SU(N+M)$ conformal theory.
     In summary, for $\Lambda\ll \Lambda_0$, i.e. at  low energy, we have $SU(M)$ gauge group that confines while at high energy $\Lambda\gg \Lambda_0$, we have a conformal field theory with two copies of $SU(N+M)$ group. Pure glue QCD, with large number of colors, confines in
	the IR and becomes conformal at the UV $-$ thus the brane setup gives rise to a QCD like gauge theory. To add flavor, we can
	add D7 branes but the overall setup does not change $-$ we still
	have UV conformal gauge theory that confines in the IR. In fact the walking RG flow in the UV due to the flavor seven-branes match up precisely with the 
	IR RG flow leading to confinement. 

Now of course the presence of anti branes will create tachyonic modes and system will be unstable. To stabilize the 
system against gravitational and RR forces, we
need to add world volume fluxes on the $D5/\bar{D}5$ branes. Alternatively, we can introduce $D7$ branes and absorb the anti-D5
branes as gauge fluxes on the D7 branes. Then a stable configuration of $D7$ branes with gauge fluxes in the presence of
coincident $D3/D5$ branes will be equivalent to
 stable configuration of  coincident $D3/D5$ branes and anti-D5 along with $D7$
branes. More details on the stabilization procedure will be discussed in \cite{Longpaper}. 

Observe that $D7$ branes also introduce fundamental matter and with world volume fluxes on Minkowski directions, they give rise to vector mesons as we shall see shortly. In summary, D7 branes play two crucial roles: they source anti-D5 charge and produce vector mesons for the four dimensional gauge theory. In principal different embeddings can be used for distinct purposes. In Figure 1, we have sketched a generic D7 embedding. Introducing various fluxes  will determine the precise embedding and in turn modify the gauge theory.     

\section{Gravity description}
When the 'tHooft coupling for the gauge theory is large, which can be achieved for example with large $M$, we can obtain a classical gravitational
description for the gauge theory arising from the above brane setup. The gravity action arises from low energy limit of type IIB
critical superstring action with localized sources given as:

\bea \label{Actions}
&&S_{\rm total}=S_{\rm SUGRA}+N_f\;S_{Dp}\nonumber\\
&&S_{\rm SUGRA}=\frac{1}{2\kappa^2_{10}}\int d^{10}x \sqrt{G}
\Bigg(R+\frac{\partial_M
\tau\partial^M\bar{\tau}}{2|{\rm
Im}\tau|^2}-\frac{|\widetilde{F}_5|^2}{4\cdot 5!}\nonumber\\
&&~~~~~~~~~ -\frac{G_3 \cdot \bar{G}_3}{12 {\rm Im}\tau}\Bigg)
+
\frac{1}{8i\kappa_{10}^2}\int \frac{C_4\wedge G_3\wedge \bar{G}_3}{{\rm Im} \tau}\nonumber\\
&&S_{Dp}=-\int d^{p+1}\sigma \;T_p\;e^{\frac{\phi(p+1)}{4}}\;\sqrt{-f}
\left(1+e^{-\phi}\frac{1}{4} \widetilde{F}^{ab}\widetilde{F}_{ab}\right)\nonumber\\
&&~~~~~~~~~ + 
\mu_p\int \left(C \wedge e^{\widetilde{F}}\right)_{p+1}
\eea
where $N_f$ is number of $Dp$ branes, $\tau=C_0+ie^{-\phi}$, $F_1=dC_0$ and $G={\rm det}g_{PQ}, P,Q=0,..,9$ with
 $g_{PQ}$ is the metric in Einstein frame. Also 
$G_3=F_3-\tau H_3$, $f={\rm det} f_{ab}$ with $f_{ab}=g_{PQ} \partial_a X^P \partial_b X^Q$. Note that  $\widetilde{F}_{ab}=F_{ab}+B_{ab}$, $F_{ab}$ is the world volume flux, $B_{ab}=B_{PQ} \partial_a X^P \partial_b X^Q$ with $B_{PQ}$ being the NS-NS two form and $\widetilde{F}_{ab}$ is raised 
or lowered with the pullback metric $f_{ab}$. The background warped metric takes
the following familiar form 
\bea\label{metric}
&&ds^2=g_{PQ}\;dx^P dx^Q\equiv g_{\mu\nu} \;dx^\mu dx^\nu+g_{mn}\; dx^m dx^n\nonumber\\
&&=-e^{2A+2B}dt^2+e^{2A}d\vec{x}^2+e^{-2A-2B}\tilde{g}_{mn} dx^m dx^n
\eea
where $d\vec{x}^2=dx^2+dy^2+dz^2$, $\mu,\nu=0,..,3,m,n=4,..,9$ and the internal unwarped metric  is given by 
$\tilde{g}_{mn}\equiv \tilde{g}_{mn}^0+\tilde{g}_{mn}^1 $. Here
 $\tilde{g}_{mn}^0$ describes the base of a deformed cone with or without resolution or squashing,  while $\tilde{g}_{mn}^1 $ is the perturbation due
to the presence of fluxes and localized sources \cite{resco}. 
A resolved-deformed cone refers to the base where cycles never go to zero size and squashing describes the deviation of shapes from an ordinary $n$-sphere. Resolution, deformation and squashing parameters are  dual description of a particular expectation value of gauge invariant combinations of bifundamental matter fields at the far IR of the gauge theory. The right figure in Figure 1 is a sketch of a warped resolved-deformed and sqaushed conifold which captures the most general dual gravity corresponding to a confining gauge theory with some expectation value of baryonic operators.  When resolution and squashing are set to zero, we also have non-zero expectation values, provided we consider deformed cone just like Klebanov-Strassler  \cite{Klebanov:2000hb}. When we are away from the resolved-deformed tip of the cone and there is no squashing, $\tilde{g}_{mn}^0$ is the metric of $R^1\times T^{1,1}$. 

The action (\ref{Actions}) in the absence of any localized sources can describe the gauge theory arising from the brane setup of Figure
1, provided  $G_3\neq 0$. When $G_3=0$, one can obtain an $AdS_5\times T^{1,1}$ geometry which describes a CFT \cite{Klebanov:1998hh}. The presence of localized sources allow us to  patch together a warped deformed conifold geometry with $G_3\neq 0$ at small radial distances to an asymptotically  
$AdS_5\times T^{1,1}$ geometry. The localized sources have to alter $G_3$, so we look for $D5$ or anti-$D5$ branes. 
We can also dissolve these branes as gauge fluxes on $D7$ branes. We take the latter approach since it is easier to find stable
$D7$ brane embeddings.      

The $D7$ branes fill up 
Minkowski space
$(t,x,y,z)$, stretching along radial $r$ direction and filling up $S^3$ inside the $T^{1,1}=S^3\times S^2$. In the absence of resolution and squashing, \cite{Mia:2013pca} proposed  D7
branes embeddings that source world volume fluxes $\widetilde{F}_2$ inducing anti-D5 charge. There were two branches of the $D7$
brane and the world volume flux on each branch modifies the background RR and NS-NS three form flux, resulting in the following
fluxes
\bea \label{G3solu1}
F_3&=&\frac{M\alpha'}{2}\omega_3+4\kappa_{10}^2\widetilde{M}N_f\alpha' \mu_7 \Bigg(F(r) \widetilde{\omega_3}^1+H(r)\widetilde{\omega_3}^2\Bigg)\nonumber\\
  H_3&=&  \frac{\ast_6\left(e^{B}F_3\right)}{{\rm Im}\tau} 
 \eea 
where $\ast_6$ is the hodge star for the metric $g_{mn}$. The definitions of the three forms 
$\omega_3,\widetilde{\omega_3}^1,\widetilde{\omega_3}^2$ and of the scalar functions $F(r),H(r)$ can be found in \cite{Mia:2013pca} (see also earlier works \cite{FEP, melting, chempot}). 
 The effective
number of $D5$ branes  in the dual gauge theory can be obtained using Gauss' law: 
\bg
M_{\rm eff}^{\rm total}=\int F_3
\nd 
For a given value of $M$, we can choose $\widetilde{M}$ such that 

\bg
\int_{r\rightarrow \infty} F_3=0~~\Rightarrow~~  M_{\rm eff}^{\rm total}(r\rightarrow \infty)=0
\nd
 Since the radial coordinate $r$ is
dual to the energy scale of the gauge theory, we find that the total D5 brane charge vanishes in the far UV and we are left only
with D3 branes. Thus For $r<r_0$, that is $\Lambda<\Lambda_0$, one finds using (\ref{G3solu1}) that
$M_{\rm eff}^{\rm total}(\Lambda<\Lambda_0)\sim M$ i.e. we have $M$ units of D5 charge. 

The introduction of $r_0$ gives rise to a scale and we divide the geometry into three regions: a classification that will be particularly useful in studying meson spectrum. 

\noindent Region I:  $r<r_0$; Region II: $r\sim r_0$; Region III: $r\gg r_0$ 

\subsection{Confinement and meson spectrum}
The form of the metric (\ref{metric}) describes a manifold $X$ with or without a black hole. When $B=0$, we have a geometry without a
black hole while in the presence of a black hole, we have a horizon with radial location $r=r_h$ such that $e^{B(r_h)}=0,
e^{2A(r_h)}\neq 0$. The temperature of the gauge theory dual to $X$ is determined by the
singularity structure of $X$ in the following way: analytically continue $\tilde{\tau}=-it$ to obtain the Euclidean metric where
$\tilde{\tau}\in [0,\beta]$. Then temperature is given by $T^{-1}=\beta$. In the presence of a black hole, $X$ is singular and
removing the singularity fixes the period $\beta$. Thus for a black hole geometry, the temperature is
 related to the horizon. On the other hand, in the absence of a black hole we pick any value of $\beta$ since we consider warp factors $e^{-4A}=h$ to be  regular on $X$. If we denote the
 `vacuum' geometry without a black hole by $X^1$ with on-shell action ${\cal S}^1$ and black hole geometry by $X^2$ with on-shell action
 ${\cal S}^2$, then at a given temperature the geometry with smaller value of on-shell action will be preferred. At $T=T_c$, $\triangle {\cal S}\equiv
 {\cal S}^2-{\cal S}^1=0$ and  we have a phase transition. At $T<T_c$, $\triangle {\cal S}>0$ and $X^1$ is preferred 
 \cite{Hawking:1982dh}\cite{{Witten:1998zw}}\cite{Mia:2013pca}\cite{Mia:2012ue}. Since there is no black
 hole, $X^1$ corresponds to zero entropy and confinement. On the other hand for $T>T_c$, $\triangle {\cal S}<0$ and black hole
 geometry is preferred. Since the black hole has non-zero entropy, the gauge theory is in the deconfined phase and $T=T_c$
 corresponds to confinement/deconfinement transition temperature.
 
 Thus at small temperatures $T<T_c$, we can consider the `vacuum' geometry without a black hole since it describes the confined phase. We
 can obtain the meson spectrum by introducing additional $D7$ branes embedded as  probes in the geometry with metric (\ref{metric}) in the limit
 $B=0$. Note that these probe $D7$ branes differ from the  $D7$ branes considered in \cite{Mia:2013pca}. The additional probe branes with world volume fluxes in Minkowski directions give rise to QCD like vector mesons. The world volume fluxes on the background $D7$ branes have no legs in the Minkowski directions and they represent dissolved anti-$D5$ branes necessary for a UV complete theory. 
 
 Before going into the details of the probe brane embedding, observe that the energy scale $\Lambda_0$ corresponding to $r_0$ provides us a notion of UV and IR energies. Since mesons appear at low energies, we expect the spectrum to be sensitive to 
 Region I with $r<r_0$ and the  characteristic mass scale for the mesons to be set by $r_c<r_0$. This mass scale manifests itself in the dual geometry via D7 embedding that stretches from $r=\infty$ to $r=r_c$. Now if we consider the trivial embedding where the pull back metric is the spacetime metric and the brane is a point in the transverse directions with embedding function being constant, then the brane will  slide down to the region $r<r_c$ due to gravitational pull. If we consider the brane to have some shape i.e. the embedding function is not a constant, then it will be possible for it to end at $r_c$, just like the U shaped embedding in \cite{Kuperstein}. However, the spectrum analysis becomes quite involved for a non-trivial embedding, since the gauge fluxes and embedding will be coupled. 
 
 One alternative to avoid such complications  is to cut off the geometry at $r=r_c$ and only consider the $r\ge r_c$ region. In this scenario, the constant embedding $D7$ brane will extend from $r=\infty$ to $r=r_c$ and $r_c$ will provide the characteristic scale of the mesons. For instance, all the meson masses will be expressed in units of $r_c$. In the following analysis, we will introduce $r_c$ as a cutoff in the geometry which essentially acts as an IR cutoff in the gauge theory. The cutoff geometry will have the same form (\ref{Actions}) as it's action with the boundary condition for metric and fluxes at $r=r_c$ consistent with the bulk solution without the cutoff.        
 
To draw a parallel with the  celebrated Sakai-Sugimoto model \cite{SS1, SS2}, we T-dualize the metric (\ref{metric})  along the $\psi$ coordinate of the conifold geometry and  analyze the DBI action of a single D6 brane. We pick world volume parametrization $(\sigma^0,..,\sigma^6)=(t,x,y,z,r,\phi_2,\theta_2)$ and  the brane is a point inside $S^3$ with the embedding: $(\theta_1, \phi_1,\psi) = (0, 0, \psi(r))$. The induced metric and $B$-field on the D6 world volume are:
\begin{align}\label{warp}
&{f}_{ab} dX^a dX^b= 
\frac{\eta_{\mu\nu}dx^\mu dx^\nu}{\sqrt{h(r)}} +\frac{9 L^4 }{r^2 \sqrt{h(r)}} {\psi'(r)}^2 dr^2\nonumber\\
&~~~~~~~ + \sqrt{h(r)} \left[dr^2+ r^2\left(\text{d$\theta $}_2^2 + (u + \sin(\theta_2)^2) \text{d$\phi $}_2^2 \right)\right] \nonumber\\
&B_2 = 3 g_s M \log (r/r_c) \sin \left(\theta _2\right) \text{d$\theta$}_2\wedge \text{d$\phi $}_2 \nonumber\\
&~~~~~~~~~ + 2 L^2 \psi'(r) \cos \left(\theta _2\right) \text{dr} \wedge \text{d$\phi $}_2 \\
&e^{-\phi(r)} = \frac{h(r)^{\frac{1}{4}} r}{6 g_s},~~h(r) = \frac{L^4  \left(1 + \frac{3 g_s M^2  \log (r/r_c)}{2 \pi  N}\right)}{4 \, r^4}\nonumber
\end{align}
where $L^4=27 \pi  N \alpha'^2$ and $u$ is a {\it squashing} parameter describing a squashed sphere at the base of the cone.
Since we will only consider $u << 1$, we do not show its dependence on the RR fields; and $(B_2, \phi)$ are kept independent of $u$. 
Also note that the warp factor $h(r)$ above is only valid for Region I and we are  
considering a region away from resolved/deformed base. Thus essentially we consider T-dual of the warped squashed $T^{1,1}$.  
Solving the embedding equation for $\psi(r)$, one finds that $\psi(r) = c \, (\text{constant})$ is a solution \cite{Longpaper}. For the study of meson 
spectrum one needs to study fluctuations of embedding for which more convenient coordinates are  $(Y,Z)$:
\begin{align}
&Y = \rho \cos(\theta),& \quad Z &= \rho \sin(\theta) \nonumber\\
&\rho = \sqrt{Y^2 + Z^2},& \theta &= \arctan\left(\frac{Z}{Y}\right)\nonumber\\
&r = r_c e^{\rho},& \quad \psi &= \frac{2 c}{\pi} \, \theta \label{EQ:rdef}
\end{align}
In this new coordinate system, the constant embedding is described by $Y=0$. Also note that the coordinate transformation makes the IR cutoff $r=r_c$ manifest since in the new coordinate $\rho\ge 0$ spans the entire cutoff geometry. Finally, if $r_c$ is bigger than the deformation parameter that appears in Klebanov-Strassler theory \cite{footnote},
%\footnote{The deformed conifold is characterized by non-zero size of $S^3$ at base of the conifold, which in turn determines the scale of confinement. The size corresponds to %expectation values of gauge invariant combinations of bifundamental fields and gives a length scale. If $r_c$ is bigger than this length scale, then the cutoff geometry can be% identified with $r>r_c$ regions of deformed cone.},  
we effectively consider mesons heavier than the confinement scale.  

\subsubsection{Vector Mesons Action}
The vector mesons arises by considering  gauge flux ($A_M$)  along the Minkowski ($t,x,y,z$) and $Z$ directions.
\begin{align}
A_M &= 
\begin{cases}
A_\mu(x^\mu,Z) \quad &\text{when $M = \mu \in \{t,x,y,z\}$} \\
A_Z(x^\mu,Z) \quad &\text{when $M = Z$} \\
0 \quad &\text{when $M \in \{\theta_2,\phi_2 \}$}
\end{cases} \nonumber\\
F_{M N} &= \partial_M A_N - \partial_N A_M
\end{align}
Looking at terms quadratic in $F_{M N}$ in the DBI action, we have:
\bea
S_{D6} &=& -T \int d^4x \, dZ \, d\theta_2 d\phi_2 \, e^{-\phi(r(0,Z))} \sqrt{-\text{det}(g_6 + {\cal B}_6)} \nonumber\\
&=& - (2 \pi \alpha')^2 T \int d^4x  dZ \Big( v_1(Z) \, \eta^{\mu \nu} F_{\mu Z} F_{\nu Z} \nonumber\\
&&~~~~~~~~~~~ + v_2(Z) \eta^{\mu \nu} \eta^{\rho \sigma} F_{\mu \rho} F_{\nu \sigma} + \ldots\Big)
\eea
where ${\cal B}_6 \equiv B_6 + 2 \pi \alpha' $; and 
$v_1(Z)$ and $v_2(Z)$ are even functions of $Z$, which also have non-trivial dependence on $u, g_s$ and $M$. The algebraic expressions of these functions can be found in \cite{Longpaper}.

We now expand $A_\mu$ and $A_Z$ in eigenmodes using two sets of eigenfunctions $\{\alpha_n(Z), n \ge 1\}$ and $\{\beta_n(Z), n \ge 0\}$
\bea
A_\mu(x^\mu,Z) &=& \sum_{n=1}^{\infty} B_{\mu}^{(n)}(x^\mu)\alpha_n(Z)\nonumber\\
A_Z(x^\mu,Z)&=& \sum_{n=0}^{\infty} \varphi^{(n)}(x^\mu)\beta_n(Z)\label{EQ:Amu}
\eea
Focusing on terms proportional to $\alpha_n^2$, we obtain terms reminiscent of the vector mesons terms of QCD.
\bea
&&S_{\alpha_n^2} = - (2 \pi \alpha')^2 T \int d^4x \, dZ \sum_{m,n}
 \big[  v_2(Z) \, F_{\mu \nu}^{(n)} F^{\mu \nu (m)} \alpha_n \alpha_m \nonumber\\
 &&~~~~~~~~~~ + v_1(Z) B_{\mu}^{(m)} B^{\mu (n)}\dot{\alpha}_m\dot{\alpha}_n \big] \label{EQ:Valpha2}
\eea
We will now impose the following conditions on $\alpha_n$,
\begin{align}
&- \partial_Z \left( v_1(Z) \, \partial_Z \alpha_n \right) = 2 \, v_2(Z) \, m_n^2 \alpha_n \label{EQ:eeq}\\
& (2 \pi \alpha')^2 T \int \, dZ \, v_2(Z) \, \alpha_m \alpha_n = \frac{1}{4}\delta_{mn} \label{EQ:alphanormcond}
\end{align}
where $m_n^2 \equiv \lambda_n \mathcal{M}^2$ is the effective squared-mass of each vector meson and $\lambda_n$ is the eigenvalue of the corresponding mode. As expected, the mass scale $\mathcal{M}^2$ is given by $\frac{r_c^2}{4 \pi N\alpha'^2}$. From the last two equations, we can derive the following identity:
\begin{align}\label{iden}
(2 \pi \alpha')^2 T \int \, dZ \, v_1(Z)  \dot{\alpha}_m \dot{\alpha}_n &= \frac{1}{2} \, m_n^2 \delta_{m n}
\end{align}
Using the above relation, the action (\ref{EQ:Valpha2}) takes the form resembling  QCD.
\begin{align}
S_{QCD} = - \sum_{n = 1}^{\infty}\int d^4x 
 \left( \frac{1}{4} F_{\mu \nu}^{(n)} F^{\mu \nu (n)}
 + \frac{1}{2} m_n^2 B_{\mu}^{(n)} B^{\mu (n)} \right) 
\end{align}
and thus $m_n$ can indeed be identified with the vector meson mass.

\subsubsection{Vector Mesons Spectrum}\label{S:VM}
We now solve the eigenvalue equation (\ref{EQ:eeq}) by using simple perturbation techniques with $\delta \equiv \frac{g_s M^2}{N}$ as the controlling parameter. We introduce some notation to write the problem in terms of a differential operator $\mathbf{H}_{\text{v}}$ acting on its eigenfunctions $\alpha_n$ \cite{Longpaper}.
\begin{align}
& (\ref{EQ:eeq}) \rightarrow  \mathbf{H}_\text{v} | \alpha_n \rangle =  \lambda_n |\alpha_n\rangle  \nonumber\\
& \mathbf{H}_{\text{v}} \equiv  - \frac{v_1(Z)}{2 \, \mathcal{M}^2 v_2(Z) } \left( \partial_Z^2 + \frac{v_1'(Z)}{v_1(Z)}  \, \partial_Z   \right)\nonumber\\
& f(Z) \equiv 4 \,(2 \pi \alpha')^2 T \, v_2(Z) \nonumber\\
& \langle \alpha_m | \alpha_n \rangle \equiv \int\limits_{\mathbb{R} \backslash \{0\}} dZ \, f(Z) \alpha_m \alpha_n = \delta_{mn}  \label{EQ:qmeeq}
\end{align}
We can now solve eq. (\ref{EQ:qmeeq}) up to first order in $\delta$ obtaining the eigenfunctions and eigenvalues \cite{Longpaper}. We impose that the eigenfunctions be normalizable so that the orthogonality condition in (\ref{EQ:qmeeq}) is satisfied. At zeroth order in $\delta$, the eigenfunctions are given in terms of Bessel's functions of the first kind.
\begin{align}
\alpha_n^{(0)}(Z) = C \, e^{-|Z|}  J_1\left(\sqrt{\lambda_n} e^{-|Z|}\right)
\end{align}
$C$ is determined by using the zeroth-order normalization condition. The eigenvalues are obtained by solving the following equations, which we expect for odd and even functions. These conditions also guarantee perfect orthonormality of the eigenfunctions:
\begin{align}
\alpha_n^{(0)}(0,\lambda_n) &= 0\quad (\text{Odd functions})\label{EQ:oddcond}\\
\partial_Z \alpha_n^{(0)}(0,\lambda_n) &= 0 \quad (\text{Even functions})\label{EQ:evencond}
\end{align}
For odd functions, we also add an extra $\text{sign}(Z)$ to make them truly odd. Using the same indexing as Sakai \& Sugimoto, the eigenfunctions are summarized as follows:
\begin{align}
&\alpha_{2n + 1}^{(0)}(Z) = C \, e^{-|Z|}  J_1\left(\sqrt{\lambda_{2n+1}} e^{-|Z|}\right) \\
&\alpha_{2n}^{(0)}(Z) = C \, \text{sgn}(Z) \, e^{-|Z|}  J_1\left(\sqrt{\lambda_{2n}} e^{-|Z|}\right)
\end{align}
Now the first-order correction to the eigenvalues of eq.(\ref{EQ:qmeeq}) is given by the well-known formula in perturbation theory and are expressed here as \cite{Longpaper}:
\begin{align}
&\lambda_n^{(1)} = \langle \alpha_n^{(0)}|\mathbf{H}_{\text{v}}^{(1)} | \alpha_n^{(0)} \rangle^{(0)} \nonumber\\
&\quad \mathbf{H}_{\text{v}}^{(1)} = \frac{3 \, e^{2 |Z|}}{2 \pi} \Big[ |Z| \partial_Z^2 - 2\, Z {\cal G}(u) \, \partial_Z \Big] \nonumber\\
&{\cal G}(u) \equiv 7 -\frac{4}{1+u} - \frac{192 u}{24 (1+ u) - \pi^2}
\end{align}
Thus the first order Hamiltonian and zeroth order eigenfunctions are sufficient to determine the eigenvalue and hence the mass up to first order in $\delta$. 

Note that the determination of mass requires us to solve (\ref{iden}) for which we need to perform an integral over all $Z$. However, due to normalizability of the eigenfunctons, the integrand contributes insignificantly for large $Z$. On the other hand, by taking $r_c$ small, $Z$ integration will be dominated by Region I.  Thus, we conclude that choosing IR cutoff $r_c$ arbitrarily small, the meson spectrum can be made independent of Region II and Region III and thus insensitive to UV modes of the gauge theory. Now of course we cannot choose $r_c$ arbitrarily small, since then we need to consider deformed, resolved cone and our analysis do not apply. However, the normalizability of eigenmodes suggest that even for reasonably large $r_c$, the mass computation will be dominated by Region I. This is not surprising since meson physics is a low energy affair and UV effects can leave the IR intact.     

\subsubsection{Mesons Identification}
We would like to verify if this effective model of large $N$ QCD shares even more similarities with the experiments 
by comparing ratios of $m_n^2$ of well-known vector mesons. In order to do so, we must first identify which kind of mesons
are present in this effective theory by looking at their behavior under charge conjugation ($\mathcal{C}$) and parity ($\mathcal{P}$). Parity operator is a Lorentz transformation flipping the space-like coordinates while charge conjugation corresponds to a flip of the $Z$ coordinate\cite{SS1}. Looking at the expansion of the four-dimensional gauge potential (\ref{EQ:Amu}), we conclude that $B_{\mu}^{(n)}$ must be odd (resp. even) under parity/charge conjugation when $\alpha_n$ is even (resp. odd) in order for $A_\mu$ to behave as a 4-vector and acquire an overall sign under charge conjugation.

Knowing the eigenvalues of each vector mesons under $\mathcal{P}$ and $\mathcal{C}$, we can identify them using the Particle Data Group (PDG) database \cite{PDG} where we use their mass measurements $M_{\text{PDG}}$ for comparison. Also, we concentrate on fields that are vectors of the approximate isospin $SU(2)$ symmetry as it was clarified in \cite{Son:2003/04}. In table \ref{vecmeson} we summarize our knowledge of each vector mesons $B_{\mu}^{(n)}$ both at zeroth and first order in
$\delta = {g_sM^2/N}$.

\begin{table}[h]
\caption{Mass ratio predictions with $0 < \delta < 0.4$ and a maximal correction of 70\%; $\delta$ = 0.4000 and $u$ = 0.0528 minimize $\chi^2/2$ to 1.4200. Results are compared with both Sakai-Sugimoto \cite{SS1} and PDG values \cite{PDG}.}
\begin{equation*}
\begin{array}{|c|c|c|c|c|c|}
\hline
 \text{} & \lambda _n/\lambda _m & \text{Sa-Su} & R_{n/m}^{(0)} & R^{(1)}_{n/m} & R_{n/m}^{\text{PDG}} \\
\hline
 m_{a_1(1260)}^2/m_{\rho (770)}^2 & \lambda _2/\lambda _1 & 2.32 & 2.54 & 2.34 & 2.52 \\
\hline
 m_{\rho (1450)}^2/m_{\rho (770)}^2 & \lambda _3/\lambda _1 & 4.22 & 5.27 & 4.19 & 3.57 \\
\hline
 m_{a_1(1640)}^2/m_{\rho (770)}^2 & \lambda _4/\lambda _1 & 6.62 & 8.51 & 6.14 & 4.51 \\
\hline
 m_{\rho (1700)}^2/m_{\rho (770)}^2 & \lambda _5/\lambda _1 & 9.53 & 12.95 & 8.46 & 4.92 \\
\hline
\end{array}
\label{vecmeson}
\end{equation*}
\end{table}

Although the results to first order in $\delta$ presented in table \ref{vecmeson} are slightly better than the ones of 
Sakai-Sugimoto \cite{SS1} and others \cite{others},  
so far the discussions have been confined to the massive KK modes of the {\it massless} open string sector of the theory. However open strings also have massive modes, and in principle these modes can also be identified as mesons. As an example, in table \ref{vecmeson}, the $1^{--}$ states $\rho(1450)$ and $\rho(1700)$ could also appear from the massive stringy sector of the model. 
For AdS space an analysis has been performed in \cite{SSvec}, where it was shown that the vector meson spectra do get contributions from the massive stringy modes. A similar analysis for our case is rather hard to perform because RR states do not decouple in the simple way as in \cite{SSvec}, rendering the quantization procedure highly non-trivial. This means a definite predictions cannot be made at this stage. Thus for our case we will continue to use the massless open string sector to study the vector mesons. 
The other three states appearing in table \ref{vecmeson}, namely $1^{--}[\rho(770)]$, $1^{++}[a_1(1260)]$ and $1^{++}[a_1(1640)]$, are only from the massless open string sector. 
In addition to that, the massless open string sector cannot be identified with scalar mesons of QCD, since a certain $Z_2$ symmetry is not shared by the theories as pointed out in \cite{SSvec}. Thus our analysis is limited to the study of vector mesons. More details on scalar meson including its spectrum  will be given in \cite{Longpaper}.  

\section{Conclusion}
In this note, we have summarized a proposal for a brane configuration and the dual geometry that can mimic several features of large $N$ QCD. For the first time using a top-down model where  classical description is sufficient, we are able to reproduce some aspects of RG flow and vector meson spectrum that are consistent with  QCD. Our study of normalizable modes on probe branes suggests that vector mesons heavier than the deconfinement scale but lighter than $\Lambda_0$ (above which the theory is almost CFT) are independent of UV ($\Lambda>\Lambda_0$) physics. 
By choosing two parameters, we can predict four mass ratios and the results show considerable improvement over previous similar approaches
 in reaching agreement with experimental data. Although there are many similarities, the brane theory cannot be identified with QCD. Rather the holographic techniques should be best utilized as tools to gain analytic understanding of the non-perturbative regimes of QCD. 

\noindent {\it Acknowledgements}: It is a great pleasure to thank Shigeki Sugimoto, Peter Ouyang and Martin Kruczenski for helpful discussions. 
The work of K. D., C. G., M. R., and O. T. is supported in part by the Natural Sciences and Engineering Research Council of Canada, and that of M. M is supported in part by 
DOE grant number ${\rm DE-SC}0007884$.


\begin{thebibliography}{150}
\bibitem{resco}
The resolved deformed cones, with or without squashing, are not Calabi-Yau manifolds but more general non-K\"ahler manifolds.
\bibitem{footnote}
The deformed conifold is characterized by non-zero size of $S^3$ at base of the conifold, which in turn determines the scale of confinement. The size corresponds to expectation values of gauge invariant combinations of bifundamental fields and gives a length scale. If $r_c$ is bigger than this length scale, then the cutoff geometry can be identified with $r>r_c$ regions of deformed cone.
\end{thebibliography}
\end{document}